# Modeling of torsion stress giant magnetoimpedance in amorphous wires with negative magnetostriction

N.A. Buznikov[*], C.O. Kim

*Research Center for Advanced Magnetic Materials, Chungnam National University,*

*220 Gung-dong, Yuseong-gu, Daejeon 305-764, Republic of Korea*

**Abstract**

A model describing the influence of torsion stress on the giant magnetoimpedance in amorphous wires with negative magnetostriction is proposed. The wire impedance is found by means of the solution of Maxwell equations together with the Landau–Lifshitz equation, assuming a simplified spatial distribution of the magnetoelastic anisotropy induced by the torsion stress. The impedance is analyzed as a function of the external magnetic field, torsion stress and frequency. It is shown that the magnetoimpedance ratio torsion dependence has an asymmetric shape, with a sharp peak at some value of the torsion stress. The calculated field and stress dependences of the impedance are in qualitative agreement with results of the experimental study of the torsion stress giant magnetoimpedance in Co-based amorphous wires.

---

[*] Corresponding author. *E-mail address:* n_buznikov@mail.ru

# 1. Introduction

The giant magnetoimpedance (GMI) effect has been studied intensively in recent years. The GMI implies a strong dependence of the impedance of a soft magnetic conductor on an external magnetic field and has been observed in different amorphous and nanocrystalline materials [1–4]. The effect is promising due to its possible use for a wide range of applications.

It has been found that Co-based amorphous wires with nearly zero negative magnetostriction provide one of the best conditions for the strong GMI effect. It is well known that internal stresses induced during the fabrication process result in a peculiar domain structure in such wires. The wire has the internal axially magnetized core and the outer shell with a circular or helical magnetic anisotropy [5–9]. Within the range of low external magnetic field, the outer shell of the wires may consist of the circular domains with opposite magnetization direction (the so-called bamboo domain structure) [5,7,8]. Due to the magnetostrictive origin of the anisotropy, the application of external stresses may lead to a change in the wire magnetic structure and large variations in the impedance.

Much attention has been paid recently to the influence of external stresses on the GMI effect in Co-based amorphous wires [10–23]. It has been found that the application of the tension stress leads to a significant change in the field dependence of the GMI ratio. The value of the external magnetic field corresponding to the maximum of the impedance increases, and the GMI ratio decreases with the growth of the tension stress [10–15,21]. It has been demonstrated also that torsion stress dependence of the GMI ratio in Co-based amorphous wires has an asymmetric character [15–19], and the current annealing results in a decrease in the asymmetry [18,19]. Moreover, it has been found that the application of only the torsion stress without any magnetic field gives rise to a change in the impedance [14,15]. The stress dependence of the magnetoimpedance response can be used for the development of highly-sensitive stress sensors [24–27].

The aim of this paper is to present a model for analysis of the torsion stress effect on the GMI in amorphous wires with negative magnetostriction. It is shown that the magnetoelastic energy contribution due to the external torsion stress leads to the changes in the magnetic structure at the wire surface. Assuming a simplified spatial distribution of the magnetoelastic anisotropy induced by the torsion stress, the expressions for the wire impedance are found by means of the solution of Maxwell equations taking into account a tensor form of the permeability. The GMI ratio has a maximum at some value of the torsion stress, which depends on the intrinsic anisotropy axis angle. The results obtained allow one to explain main features of the torsion stress dependence of the GMI effect in Co-based amorphous wires observed in experiments.

## 2. Model

Let us consider an amorphous negative magnetostrictive wire of length $l$ and radius $r$ submitted to a homogeneous torsion stress $\xi$. The AC current $I=I_0\exp(-i\omega t)$ flows along the wire (along $z$-axis), and the external magnetic field $H_e$ is parallel to the current. Further, we ignore for simplicity the effect of the longitudinally magnetized inner core and assume that the whole wire has a helical magnetic anisotropy, and the anisotropy axis makes the angle $\psi$ with the circular direction.

It is assumed also that the wire has a single-domain structure, and the variations in the magnetization are related to the magnetization rotation process. The later approximation is known to be valid for not too low current frequencies [3]. The equilibrium magnetization angle $\theta$ with respect to the circular direction can be found by the free energy minimization. The free energy density $U$ consists of the anisotropy term, Zeeman energy and the magnetoelastic energy arising due to the application of the torsion stress [28,29]:

$$U = (MH_a/2)\sin^2(\theta-\psi) - MH_e\sin\theta + (MH_\xi/2)\sin^2(\theta-\pi/4). \tag{1}$$

Here $M$ is the saturation magnetization, $H_a$ is the anisotropy field and the last term represents the helical anisotropy induced by the torsion stress [9], where

$$H_\xi = 3\lambda_s\Gamma\xi\rho/M = H_{\max}\rho/r, \tag{2}$$

$\lambda_s < 0$ is the magnetostriction constant, $\Gamma$ is the shear modulus, $\rho$ is the radial coordinate and $H_{\max} = H_\xi(r) = 3\lambda_s\Gamma\xi r/M$ is the value of the magnetoelastic anisotropy field $H_\xi$ at the wire surface. As follows from Eq. (2), for positive stress (clockwise rotation) we have $H_\xi < 0$, and for negative $\xi$ (counter-clockwise rotation) the magnetoelastic anisotropy field $H_\xi$ is positive. Note that the magnetoelastic anisotropy induced by the torsion stress is non-uniform and depends on the radial coordinate. This makes the modeling of the effect of the torsion stress on the GMI rather difficult.

The minimization of the free energy density gives the following equation for the equilibrium magnetization angle $\theta$:

$$H_a \sin(\theta-\psi)\cos(\theta-\psi) - (H_\xi/2)\cos 2\theta - H_e \cos\theta = 0. \qquad (3)$$

Fig. 1 shows the magnetization curves calculated by means of Eq. (3) for different values of $H_\xi$. At low external field $H_e$, there are two solutions of Eq. (3) with different equilibrium magnetization angles. Within this field range, the bamboo domain structure may appear in the wire. The application of the torsion stress changes drastically the magnetization distribution. In the absence of the stress and at low negative $H_\xi$, the circular component of the magnetization has the positive sign at $H_e > H_a$, whereas for higher negative $H_\xi$, the circular magnetization component is negative (see Fig. 1(b)). Since the magnetoelastic anisotropy induced by the torsion stress increases linearly with the radial coordinate, this means that the region with the negative sign of the circular magnetization component appears at the wire surface at sufficiently high $\xi$.

It can be readily shown that the changes in the sign of the circular magnetization component takes place when $H_\xi = -H_a \sin 2\psi$. Taking into account the radial distribution of $H_\xi$, we obtain the threshold value of the torsion stress:

$$\xi_{cr} = -MH_a \sin 2\psi / 3\lambda_s \Gamma r. \qquad (4)$$

To calculate the GMI response, we suppose a simplified spatial distribution of the field $H_\xi$, which allows one to find analytical expressions for the impedance. Let us consider at first the case of $\xi < \xi_{cr}$. It is assumed that the field $H_\xi$ is uniform over the wire and equals its maximum value at the wire surface, $H_\xi = H_{max}$. In the approximation of a local relationship between the magnetic field and the magnetization, the wire impedance can be found by means of the solution of Maxwell equations together with the Landau–Lifshitz equation. In the case of the strong skin effect, the impedance $Z$ is given by the following expression [30–32]:

$$Z/R_{DC} = (1-i)(r/2\delta) \times [(\mu+1)^{1/2}\sin^2\theta + \cos^2\theta]. \tag{5}$$

Here $R_{DC} = l/\pi\sigma r^2$ is the DC wire resistance, $\sigma$ is the wire conductivity, $\delta = c/(2\pi\sigma\omega)^{1/2}$ is the skin depth in non-magnetic material, $c$ is the velocity of light, the equilibrium magnetization angle $\theta$ is given by Eq. (3) and the effective permeability $\mu$ can be found from a solution of the linearized Landau–Lifshitz equation. Using the standard procedure [30,32] and taking into account the effect of $H_\xi$, we have for $\mu$

$$\begin{aligned}\mu &= \frac{\gamma 4\pi M}{\omega_1 - i\kappa\omega - \omega^2/(\omega_2 - i\kappa\omega)}, \\ \omega_1 &= \gamma[H_a\cos\{2(\theta-\psi)\} + H_\xi\sin 2\theta + H_e\sin\theta], \\ \omega_2 &= \gamma[4\pi M + H_a\cos^2(\theta-\psi) + H_\xi\cos^2(\theta-\pi/4) + H_e\sin\theta].\end{aligned} \tag{6}$$

Here $\gamma$ is the gyromagnetic constant and $\kappa$ is the Gilbert damping parameter.

In the case of low frequencies, when the skin effect is negligible, the expression for the impedance can be found by means of the asymptotic-series expansion of Maxwell equations, which gives [32]

$$Z/R_{DC} = (kr/2)J_0(kr)/J_1(kr) + 3(r/3\delta)^4\mu^2\sin^2\theta\cos^2\theta, \tag{7}$$

where $J_0$ and $J_1$ are the Bessel functions of the first kind and $k = (1+i)(1+\mu\sin^2\theta)^{1/2}/\delta$.

In order to take into account the changes in the magnetization distribution near the wire surface at $\xi > \xi_{cr}$, let us subdivide the wire into two regions. In the inner region, $\rho < \rho_c$, the circular magnetization component satisfying Eq. (3) has the positive sign, $\cos\theta_1 > 0$. It is assumed for simplicity that the magnetoelastic anisotropy field $H_{\xi 1}$ equals the averaged value over this region, $H_{\xi 1} = (2/3)H_{max}(\rho_c/r)$. In the outer region, $\rho > \rho_c$, the circular magnetization component is negative, $\cos\theta_2 < 0$, and the magnetoelastic anisotropy field within this region is assumed to be equal to its maximum value at the wire surface, $H_{\xi 2} = H_{max}$. The position of the boundary between two regions, $\rho = \rho_c$, is determined by the condition $H_\xi(\rho_c) = -H_a\sin 2\psi$, which leads to

$$\rho_c = -MH_a\sin 2\psi / 3\lambda_s\Gamma\xi. \tag{8}$$

Taking into account that the fields depend only on the radial coordinate, Maxwell equations for both the regions can be reduced to two coupled differential equations for the circular and longitudinal components of the magnetic field [30–32]:

$$\frac{\partial^2 h_\varphi^{(j)}}{\partial \rho^2} + \frac{1}{\rho}\frac{\partial h_\varphi^{(j)}}{\partial \rho} - \frac{h_\varphi^{(j)}}{\rho^2} + \frac{2\mathrm{i}}{\delta^2} \times (1 + \mu_j \sin^2\theta_j) h_\varphi^{(j)} = \frac{2\mathrm{i}}{\delta^2} \times \mu_j h_z^{(j)} \sin\theta_j \cos\theta_j ,$$

$$\frac{\partial^2 h_z^{(j)}}{\partial \rho^2} + \frac{1}{\rho}\frac{\partial h_z^{(j)}}{\partial \rho} + \frac{2\mathrm{i}}{\delta^2} \times (1 + \mu_j \cos^2\theta_j) h_z^{(j)} = \frac{2\mathrm{i}}{\delta^2} \times \mu_j h_\varphi^{(j)} \sin\theta_j \cos\theta_j ,$$

(9)

where $j = 1, 2$ corresponds to the inner and the outer regions, respectively, subscripts $\varphi$ and $z$ denote the circular and longitudinal components of the magnetic field and $\mu_j$ are given by Eq. (6) at $\theta = \theta_j$ and $H_\xi = H_{\xi j}$. In the case of high enough frequency, when the effective skin depth is much less than the wire radius, the solution of Eqs. (9) inside the inner region, $\rho < \rho_c$, can be presented in the following form [30,31]:

$$h_\varphi^{(1)}(\rho) = A_1 \cos\theta_1 \exp\{k_0(\rho - \rho_c)\} + A_2 \sin\theta_1 \exp\{k_1(\rho - \rho_c)\} ,$$

$$h_z^{(1)}(\rho) = A_1 \sin\theta_1 \exp\{k_0(\rho - \rho_c)\} - A_2 \cos\theta_1 \exp\{k_1(\rho - \rho_c)\} ,$$

(10)

where $A_1$ and $A_2$ are the constants, $k_0 = (1-\mathrm{i})/\delta$ and $k_1 = (1-\mathrm{i})(\mu_1 + 1)^{1/2}/\delta$.

In the outer region, $\rho > \rho_c$, the expressions for the circular and longitudinal components of the magnetic field are given by [33]

$$h_\varphi^{(2)}(\rho) = \cos\theta_2 [B_1 \exp\{k_0(\rho - r)\} + B_2 \exp\{-k_0(\rho - r)\}]$$
$$+ \sin\theta_2 [B_3 \exp\{k_2(\rho - r)\} + B_4 \exp\{-k_2(\rho - r)\}] ,$$

$$h_z^{(2)}(\rho) = \sin\theta_2 [B_1 \exp\{k_0(\rho - r)\} + B_2 \exp\{-k_0(\rho - r)\}]$$
$$- \cos\theta_2 [B_3 \exp\{k_2(\rho - r)\} + B_4 \exp\{-k_2(\rho - r)\}] .$$

(11)

Here $B_1$, $B_2$, $B_3$ and $B_4$ are the constants and $k_2 = (1-\mathrm{i})(\mu_2 + 1)^{1/2}/\delta$. Note that in the case of low frequencies, the distribution of the fields inside the wire can be found in the form of series [30,33].

The six constants in Eqs. (10) and (11) can be obtained from the boundary conditions. First, the circular and longitudinal components of the magnetic field and their spatial derivatives should satisfy the continuity conditions at the interface between two regions, $\rho = \rho_c$:

$$h_\varphi^{(1)}(\rho_c) = h_\varphi^{(2)}(\rho_c), \quad h_z^{(1)}(\rho_c) = h_z^{(2)}(\rho_c),$$

$$(\partial h_\varphi^{(1)}/\partial \rho)\big|_{\rho=\rho_c} = (\partial h_\varphi^{(2)}/\partial \rho)\big|_{\rho=\rho_c} ,$$

$$(\partial h_z^{(1)}/\partial \rho)\big|_{\rho=\rho_c} = (\partial h_z^{(2)}/\partial \rho)\big|_{\rho=\rho_c} .$$

(12)

Furthermore, the components of the magnetic field at the wire surface are determined by the excitation conditions:

$$h_\varphi^{(2)}(r) = 2I_0/cr,$$
$$h_z^{(2)}(r) = 0.$$
(13)

Thus, at $\xi > \xi_{cr}$ the magnetic field distribution inside the wire can be found by means of Eqs. (10)–(13). The wire impedance $Z$ can be calculated as [30,32]

$$Z = \frac{2l}{cr} \times \frac{c}{4\pi\sigma} \times \left.\frac{\partial h_\varphi^{(2)}/\partial \rho}{h_\varphi^{(2)}}\right|_{\rho=r}.$$
(14)

Using Eqs. (11), we can rewrite the expression for the impedance in the following form:

$$Z/R_{DC} = \frac{r}{2} \times \frac{k_0 \cos\theta_2(B_1 - B_2) + k_2 \sin\theta_2(B_3 - B_4)}{\cos\theta_2(B_1 + B_2) + \sin\theta_2(B_3 + B_4)}.$$
(15)

## 3. Results and discussion

The calculated field dependence of the impedance is shown in Fig. 2 for different values of the applied torsion stress. The results are presented only for the range of the positive external field, since the calculated curves are symmetrical with respect to the sign of $H_e$. It follows from Fig. 2 that the impedance is very sensitive to the torsion stress. At low $\xi$, the impedance ratio increases sharply and the field corresponding to the impedance maximum decreases with the growth of $\xi$. The field sensitivity of the impedance attains its maximum at $\xi = \xi_{cr}$. At $\xi > \xi_{cr}$, the effective permeability drops, and the impedance ratio decreases monotonically (see Fig. 2). Note that the similar dependence of the impedance on the torsion stress has been observed in experiments with Co-based amorphous wires [18].

Fig. 3 presents the frequency dependence of the impedance ratio $\Delta Z$ for different values of $\xi$. This ratio is defined as the difference between the peak impedance value, $Z_{max}$, and the impedance at zero magnetic field, $\Delta Z = Z_{max} - Z(0)$. It is seen from Fig. 3 that at low and high values of the torsion stress, $\Delta Z$ increases monotonically with the frequency, whereas at $\xi \cong \xi_{cr}$, the impedance ratio is almost constant at high frequencies. Let us introduce the stress sensitivity of the impedance $\Delta Z_\xi$, which is defined as $\Delta Z_\xi = \Delta Z(\xi = \xi_{cr}) - \Delta Z(\xi = 0)$. The frequency dependence of $\Delta Z_\xi$ is shown in Fig. 4 for different values of the anisotropy axis angle $\psi$. The stress sensitivity of the impedance attains its maximum at some frequency, and the maximum shifts to higher frequencies with the increase of the anisotropy axis angle.

The impedance ratio $\Delta Z$ is shown in Fig. 5 as a function of the torsion stress at different values of the anisotropy axis angle $\psi$. It follows from Fig. 5 that the impedance ratio has the asymmetric dependence on the torsion stress with a sharp maximum at $\xi = \xi_{cr}$. The maximum shifts towards the positive stress with the increase of the anisotropy axis deviation angle from the circular direction. Similar asymmetric behavior of the GMI stress dependence has been observed in experiments with Co-based amorphous wires [15,16,18,19]. It has been demonstrated that the torsion-stress annealing increases the asymmetry, whereas the current annealing leads to more symmetrical shape of the dependence [19]. The current annealing results in a relaxation of the internal stresses induced by the fabrication process and decreases the anisotropy axis angle. On the other hand, the torsion-stress annealing develops more pronounced helical anisotropy in the wire [16]. Therefore, the calculated results allow one to explain qualitatively the experimentally observed evolution of the stress dependence of the GMI ratio related to the changes in the anisotropy axis angle.

In the model proposed, we neglect the influence of the longitudinally magnetized core on the GMI. Although the shell gives the main contribution to the GMI at sufficiently high frequencies, it has been shown that the effect of the core on the impedance may be essential at low external fields [34,35]. The corresponding modifications taking into account the longitudinally magnetized core can be made in the framework of the present approach by considering an additional region with the longitudinal anisotropy near the wire axis, and results of calculations demonstrate that the influence of the core on the torsion stress dependence of the GMI is insignificant, if the anisotropy fields in the core and in the shell have the same order of the magnitude.

It is assumed above that the permeability is determined by the magnetization rotation. This approximation is valid at sufficiently high frequencies, when the domain-walls motion is damped by eddy currents [2,3]. At low frequencies, the field dependence of the impedance exhibits the single-peak behavior due to the effect of the domain-walls motion. The contribution of the domain-walls motion to the effective permeability and GMI response at low frequencies can be found by the methods described in Refs. [8,36–38]. The relaxation frequency for the domain-walls motion is inversely proportional to the static domain-wall susceptibility, the wire diameter and the domain size [8]. Simple estimations show that for the domain width of 10 μm and the static susceptibility of the order of $10^3$, the relaxation frequency is less than 100 kHz for the wires of diameter 120 μm. Therefore, we can conclude that the contribution from the domain-walls motion to the permeability is insignificant to describe the torsion stress effect on the GMI in thick amorphous wires at $f > 100$ kHz.

Another restriction of the model is related to the assumption on simplified radial distribution of the magnetoelastic anisotropy induced by the torsion stress. It is evident that this approximation is valid in the case of the strong skin effect, when the influence of the torsion stress on the GMI response is determined by the surface layer, and the details of the spatial distribution of the field $H_\xi$ are not essential. At low frequencies, the spatial distribution of the magnetoelastic anisotropy field $H_\xi$ and the corresponding changes in the permeability with the radial coordinate should be taken into account. The model describing the influence of the real radial distribution of the magnetoelastic anisotropy on the torsion stress GMI is presently under development.

In conclusion of this section, it should be noted that we consider the case of low current amplitudes, when the voltage response is linear, and the impedance is independent of the current amplitude. At higher current amplitudes, the relation between the magnetization and the current amplitude becomes nonlinear. As a result, the second harmonic component appears in the voltage response, which can be ascribed to the asymmetry in the circular hysteresis loop at high current amplitudes. The rotational model proposed in Refs. [28,29] predicts the increase of the second harmonic amplitude with the torsion stress due to the growth of the asymmetry in the magnetization reversal process, and this prediction has been confirmed experimentally.

## 4. Conclusions

The effect of the torsion stress on the GMI in amorphous wires with negative magnetostriction is related to a competition between the magnetoelastic anisotropy induced by the torsion stress and the helical anisotropy arising from the internal stresses. It is shown that the application of the torsion stress exceeding some threshold value of $\xi_{cr}$ results in the changes in the magnetic structure at the wire surface. In the framework of the single-domain approximation, the distribution of the fields inside the wire and the wire impedance are found by means of the solution of Maxwell equations together with the Landau–Lifshitz equation. The stress dependence of the GMI ratio is the asymmetric one, with a sharp peak at $\xi = \xi_{cr}$. The value of $\xi_{cr}$ depends on the anisotropy field and its angle with respect to the circular direction, what may be used for estimations of the intrinsic anisotropy from the measurements of the GMI effect in the presence of the torsion stress. The results obtained describe qualitatively the evolution of the stress dependence of the GMI ratio with the current annealing and the torsion-stress annealing observed in the experiments with Co-based amorphous wires [15,16,18,19].

**Acknowledgment**

This work was supported by the Korea Science and Engineering Foundation through ReCAMM.

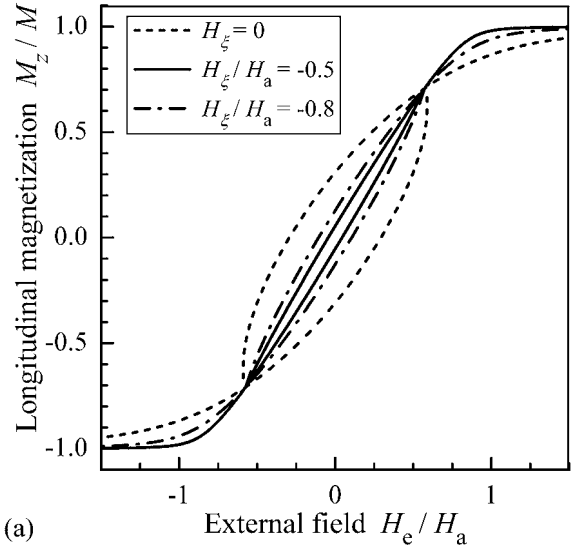

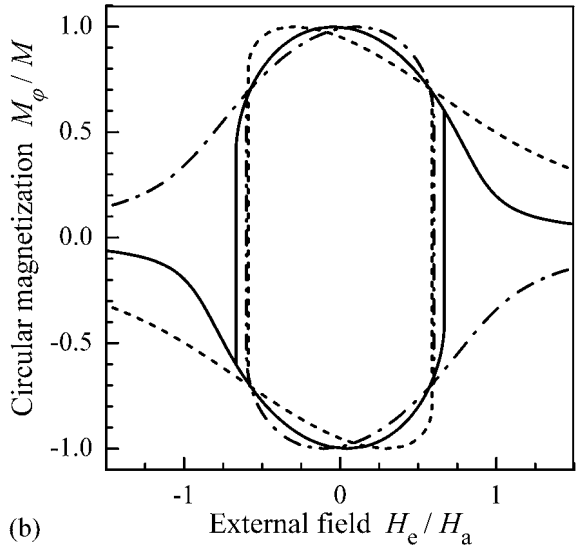

Fig. 1. Dependences of the longitudinal $M_z = M\sin\theta$ (a) and circular $M_\varphi = M\cos\theta$ (b) magnetization components on the external field $H_e$ at $\psi = 0.1\pi$ and different values of the magnetoelastic anisotropy field $H_\xi$.

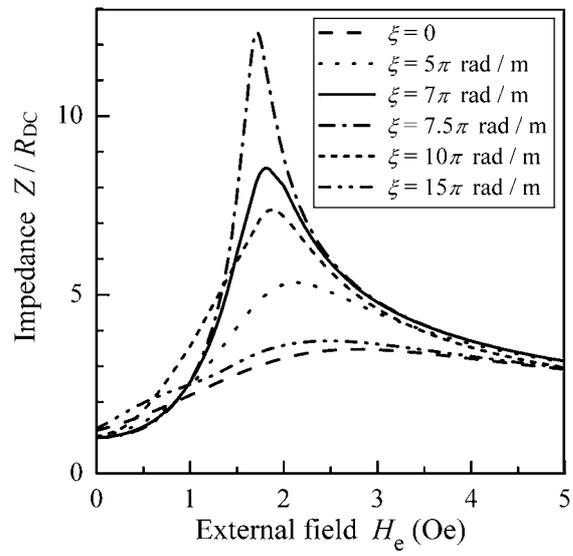

Fig. 2. The impedance $Z$ as a function of the external field $H_e$ at $f = \omega/2\pi = 500$ kHz and different torsion stress $\xi$. Parameters used for calculations are $r = 60$ μm, $M = 600$ G, $H_a = 2$ Oe, $\psi = 0.1\pi$, $\sigma = 10^{16}$ s$^{-1}$, $\kappa = 0.1$, $\lambda_s = -2 \times 10^{-7}$, $\Gamma = 80$ GPa.

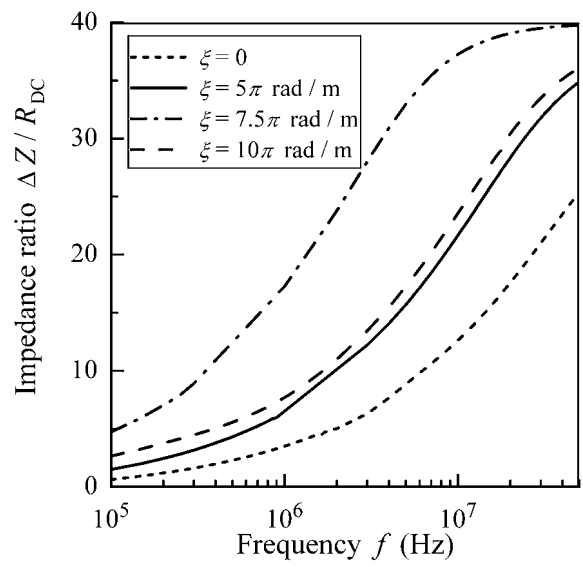

Fig. 3. Impedance ratio $\Delta Z$ versus frequency $f$ at different torsion stress $\xi$. Parameters used for calculations are the same as in Fig. 2.

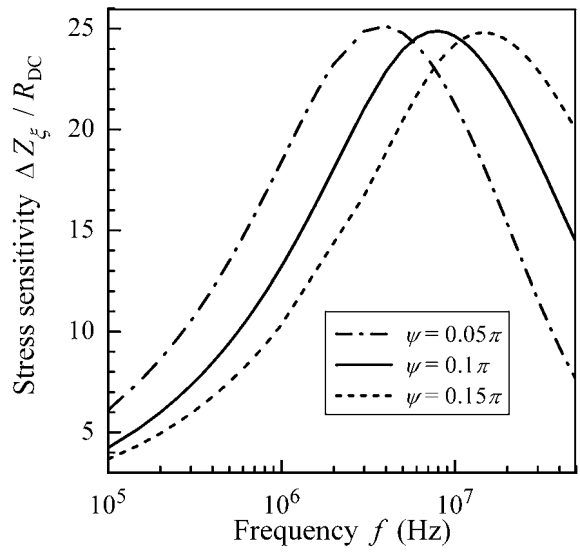

Fig. 4. Stress sensitivity of the impedance $\Delta Z_\xi$ versus frequency $f$ at different anisotropy axis angle $\psi$. Parameters used for calculations are the same as in Fig. 2.

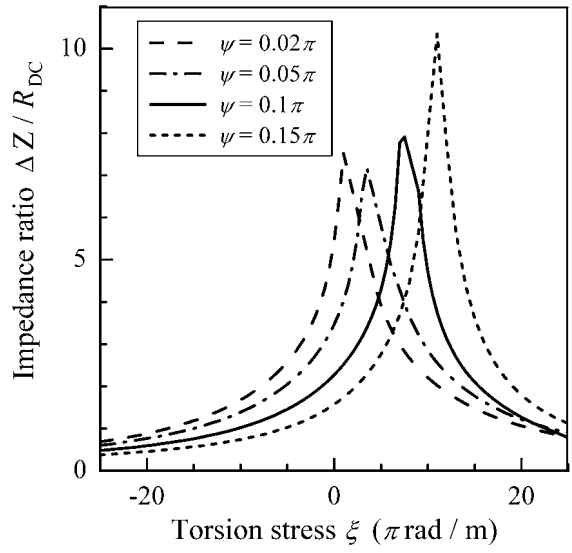

Fig. 5. Impedance ratio $\Delta Z$ versus torsion stress $\xi$ at $f=500$ kHz and different anisotropy axis angle $\psi$. Parameters used for calculations are the same as in Fig. 2.